\documentclass[conference, 10pt]{IEEEtran}

\usepackage{multicol, latexsym, amsmath, amssymb}
\usepackage{amssymb}
\usepackage{amsmath}
\usepackage{amsthm}
\usepackage{cite}
\usepackage{algorithm}
\usepackage{algorithmic}
\usepackage{romannum}
\usepackage[dvips]{graphicx}
\usepackage{stfloats}
\usepackage{comment}
\usepackage{latexsym,epsf,times,amsmath,color,amssymb,indentfirst,subfigure,fancyhdr,colortbl,bm,caption}
\usepackage{xspace}
\usepackage{graphicx}
\usepackage{bbm}
\usepackage{graphicx}
\usepackage[dvipsnames]{xcolor}
\usepackage{cuted}

\setbox1=\hbox{$\bullet$}
\usepackage{stfloats}
\ifCLASSINFOpdf

\else

\fi

\makeatletter
\newcommand*{\rom}[1]{\expandafter\@slowromancap\romannumeral #1@}
\makeatother

\DeclareMathOperator*{\argmin}{argmin} 

\begin{document}

\pagenumbering{arabic}
%
\title{Minimization of Sum Inverse Energy Efficiency for Multiple Base Station Systems}

\author{\IEEEauthorblockN{Zijian Wang\dag, Luc Vandendorpe\dag, Mateen Ashraf\dag, Yuting Mou{*}, and Nafiseh Janatian\dag}
\IEEEauthorblockA{\dag Institute of Information and Communication Technologies, Electronics and Applied Mathematics\\Universit\'{e} catholique de Louvain,
	Louvain-la-Neuve, Belgium\\{*} Center for Operations Research and Econometrics, Universit\'{e} catholique de Louvain,
	Louvain-la-Neuve, Belgium\\Emails: \{first name. last name\}@uclouvain.be
}
}

\maketitle

\newtheorem{theorem}{Theorem}
\newtheorem{lemma}{Lemma}
\newtheorem{proposition}{Proposition}
\newtheorem{corollary}{Corollary}

\begin{abstract}
A sum inverse energy efficiency (SIEE) minimization problem is solved. Compared with conventional sum energy efficiency (EE) maximization problems, minimizing SIEE achieves a better fairness. The paper begins by proposing a framework for solving sum-fraction minimization (SFMin) problems, then uses a novel transform to solve the SIEE minimization problem in a multiple base station (BS) system. After the reformulation into a multi-convex problem, the alternating direction method of multipliers (ADMM) is used to further simplify the problem. Numerical results confirm the efficiency of the transform and the fairness improvement of the SIEE minimization. Simulation results show that the algorithm convergences fast and the ADMM method is efficient. 
\end{abstract}

\IEEEpeerreviewmaketitle
\section{Introduction and Motivation}
Nowadays, energy efficiency (EE) for wireless communications is becoming a main economical and societal challenge \cite{ee}.
EE maximization is a fractional programming problem which is typically solved by Dinkelbach's algorithm \cite{bio1,bio2}. 

However, most of the works on EE are single-fraction problems.
In the literature, solving a sum-fraction problem is far more difficult than the single-fraction one.
For multiple-fraction problems, some  specific forms (e.g., the max-min problem) were studied in \cite{bio5}. A sum-fraction problem is shown to be NP-hard \cite{bio6}. The methods for finding its global optimum are quite time demanding (e.g., using branch-and-bound search \cite{bio7,bio8,bio9}). 

To find stationary-point solutions of the sum EE maximization problem, successive convex approximation methods were used in paper \cite{markku} and a Lagrangian update approach was used in paper \cite{shiwen}.
In paper \cite{yuwei}, the authors proposed a quadratic transform to reformulate the sum-fraction problem into a bi-concave one. This method decouples the numerators and denominators by introducing only one variable vector. The resulted expression with seperate numerators and denominators is always more tractable for further analysis.

Inspired by the method in \cite{yuwei}, in this paper, we propose another transform to deal with the sum-fraction minimization (SFMin) problem.
As in \cite{wangtao} and \cite{zijian}, one may aim to minimize the sum of inverse EE for more tractable expressions or analysis. Our considered problem in this paper is to minimize the sum of inverse EE for a multiple base station (BS) system.

In fact, the sum-of-inverse minimization (SIMin) leads to more fairness than the sum maximization (SMax). For better understanding this, let us take the example of maximizing the sum of two variables, as 
\begin{equation}\label{fair1}
\max_{x,y} f_1(x,y)=x+y,
\end{equation} 
and minimizing their sum-of-inverse as 
\begin{equation}
\min_{x,y} f_2(x,y)=\frac{1}{x}+\frac{1}{y}.
\end{equation}

On the one hand, it is always true that 
$
\frac{1}{2} f_1(x,y)\geq \left(\frac{1}{2} f_2(x,y)\right)^{-1},
$
which can be interpreted as the fact that the inverse of the mean of inverses is a lower bound of the mean. The bound is tight when $x=y$. 

On the other hand,
denote the solutions of the two above problems as $x_1,y_1$ and $x_2,y_2$ respectively. We have
$
x_1+y_1\geq x_2+y_2
$
and
$
\frac{1}{x_2}+\frac{1}{y_2}\leq \frac{1}{x_1}+\frac{1}{y_1}.
$
After some manipulations, we have
$
\frac{x_1}{y_1}+\frac{y_1}{x_1}\geq \frac{x_2}{y_2}+\frac{y_2}{x_2}.
$
Without loss of generality, assuming $x_1\geq y_1$ and $x_2\geq y_2$, we have 
$
\frac{x_1}{y_1}\geq \frac{x_2}{y_2}, 
$
which implies the minimization problem obtains more fairness. Note that it does not mean that $x_2=y_2$, which is full fairness. Thus, it achieves a tradeoff between fairness and overall performance.

For the scenarios where the number of terms is larger than 2, it is no longer true that 
SIMin always has a better fairness. However, with a small number of terms (e.g. less than $15$), it is true with a very high probability. We will illustrate this later in the numerical results.

In \cite{yuwei}, a quadratic transform is used to solve the sum-fraction maximization problem. Although the difference between this work and our work is only maximization and minimization, the proposed transforms are quite different.

The contributions of this paper are as follows:
\begin{itemize}
	\item A sum  energy-per-rate minimization problem is studied, which, to the best of the authors' knowledge, has never been investigated in the literature. This problem has a better tradeoff between energy efficiency and fairness concern, which is a major difference from the sum rate-per-energy maximization problem. A particular advantage is that no user is inactive by considering this problem.
	\item A novel method for solving the SFMin problem is proposed. The method decouples the numerators and the denominators, which makes it possible to optimize the numerator part and denominator part separately by using the alternating direction method of multipliers (ADMM) method. This is a general framework which can be used in other practical problems.
	\item A closed-form solution is found by the Karush–Kuhn–Tucker (KKT) conditions, which gives more insight on the solution. The closed-form solution is due to the method we proposed, which decouples the numerators and the denominators. 
\end{itemize}

\section{General models}
In this section, we begin by introducing a framework with a general optimization problem, which will be used later in a particular system model in this paper.

Let us consider an SFMin problem expressed as:
\begin{align}
\min_{\mathbf{x}\in C_x} \sum_{i=1}^I \frac{B_i(\mathbf{x})}{A_i(\mathbf{x})},
\end{align}
where $I$ is the number of terms, $\mathbf{x}$ is the variable vector whose domain is $C_x$. $A_i(\mathbf{x})$ and $B_i(\mathbf{x})$ are functions of $\mathbf{x}$, always with positive values.

This SFMin problem cannot be solved by conventional Dinkelbach's algorithm which is often used in fractional optimization. We propose a fraction transform to solve this problem. We name it 'fraction transform' because there exist fraction terms. As can be seen later in this paper, this method enables to use the ADMM to implement the optimization distributedly and obtain a closed-form solution for each subproblem.

In the following theorem, we show that it has an equivalent problem, that is,
\begin{align}\label{proposed}
\min_{\mathbf{x}\in C_x,\mathbf{t}\in \mathbb{R}_{>0}^I} \sum_{i=1}^I t_i B_i(\mathbf{x})^2+\sum_{i=1}^I \frac{1}{4 t_i}\frac{1}{A_i(\mathbf{x})^2},
\end{align}
where $\mathbf{t}$ is a newly introduced vector.

\begin{theorem}\label{theorem1}
	The solution of the minimization problem
	\begin{align}\label{proof2}
	\min_{\mathbf{x}\in C_x} \sum_{i=1}^I \frac{B_i(\mathbf{x})}{A_i(\mathbf{x})},
	\end{align}
	where $A_i(\mathbf{x})$ and $B_i(\mathbf{x})$ are positive, 
	is the same as 
	\begin{align}\label{proof3}
	\min_{\mathbf{x}\in C_x,\mathbf{t}\in \mathbb{R}_{>0}^I} \sum_{i=1}^I t_i B_i(\mathbf{x})^2+\sum_{i=1}^I \frac{1}{4 t_i}\frac{1}{A_i(\mathbf{x})^2}.
	\end{align}

\begin{proof}
	The following equation is always true: 
	\begin{multline}\label{proof1}
	t_i B_i(\mathbf{x})^2+\frac{1}{4 t_i}\frac{1}{A_i(\mathbf{x})^2}\\=\left(\sqrt{t_i} B_i(\mathbf{x})-\frac{1}{2 \sqrt{t_i}}\frac{1}{A_i(\mathbf{x})}\right)^2+\frac{B_i(\mathbf{x})}{A_i(\mathbf{x})}.
	\end{multline}
	Obviously, the optimal $\mathbf{x},\mathbf{t}$ that minimize the left-hand side of \eqref{proof1} always minimize its right-hand side. This $\mathbf{x}$ also minimizes $\frac{B_i(\mathbf{x})}{A_i(\mathbf{x})}$ because $\mathbf{t}$ can always be adapted to 
	\begin{equation}
	t_i={\frac{1}{2A_i(\mathbf{x})B_i(\mathbf{x})}}
	\end{equation}
	to force the square term in \eqref{proof1} to be zero.
	
	Therefore, the solution for \eqref{proof2} (which is $\mathbf{x}$) is always part of the solutions of \eqref{proof3} (which is $\mathbf{x},\mathbf{t}$).
\end{proof}
\end{theorem}
As an intuition, vector $\mathbf{t}$ acts as the variable $\mu$ in $A(\mathbf{x})-\mu B(\mathbf{x})$ in Dinkelbach's algorithm, where maximization of $\frac{A(\mathbf{x})}{B(\mathbf{x})}$ is assumed, to change the priorities of numerators and denominators.

Even though the numerators and the denominators are decoupled, if the problem in \eqref{proposed} is not convex, it is still difficult to solve.
In the following theorem, the convexity of the problem in \eqref{proposed} under some condition is proved.
\begin{theorem}\label{theorem2}
	If $A_i(\mathbf{x})$ is concave and $B_i(\mathbf{x})$ is convex, then the problem in \eqref{proposed} is convex for given $\mathbf{t}$.
	\begin{proof}
		We will prove the convexity by proving its Hessian matrix is positive semidefinite \cite{boyd}.
		
		The Hessian matrix of ${A}_i(\mathbf{x})^{-2}$ is 
		\begin{equation}
		\mathbf{H}_{{A}_i(\mathbf{x})^{-2}}=\frac{6}{A_i(\mathbf{x})^4}\nabla A_i(\mathbf{x})\nabla A_i(\mathbf{x})^T-\frac{2}{A_i(\mathbf{x})^3}\mathbf{H}_{A_i(\mathbf{x})}.
		\end{equation}
		Because $A_i(\mathbf{x})$ is positive and $\mathbf{H}_{A_i(\mathbf{x})}$ is negative semidefinite, $\mathbf{H}_{{A}_i(\mathbf{x})^{-2}}$ is positive semidefinite. 
		
		The Hessian matrix of ${B}_i(\mathbf{x})^2$ is 
		\begin{equation}
		\mathbf{H}_{{B}_i(\mathbf{x})^2}=2{B}_i(\mathbf{x}) \mathbf{H}_{B_i(\mathbf{x})}+2\nabla B_i(\mathbf{x})\nabla B_i(\mathbf{x})^T.
		\end{equation}
		Because $B_i(\mathbf{x})$ is positive and $\mathbf{H}_{B_i(\mathbf{x})}$ is positive semidefinite, $\mathbf{H}_{{B}_i(\mathbf{x})^2}$ is positive semidefinite. 
		
		Therefore, the objective funtion in \eqref{proposed} for given $\mathbf{t}$ is convex because its Hessian matrix is positive semidefinite.
	\end{proof}
\end{theorem}

From the analysis above, the problem can be solved in an alternating manner. 
The following convex problem is solved for a given $\mathbf{t}$:
\begin{equation}\label{pro1}
\min_{\mathbf{x}\in C_x} F(\mathbf{t})=\sum_{i=1}^I t_i B_i(\mathbf{x})^2+\sum_{i=1}^I \frac{1}{4 t_i}\frac{1}{A_i(\mathbf{x})^2},
\end{equation}
and then
\begin{equation}
t_i={\frac{1}{2A_i(\mathbf{x})B_i(\mathbf{x})}}
\end{equation}
is updated.

This proposed fraction transform enables the ADMM method for solving \eqref{pro1}, for example, as 
\begin{align}\label{admm}
\min_{\mathbf{x}\in C_x, \mathbf{z}} &\quad \sum_{i=1}^I t_i B_i(\mathbf{x})^2+\sum_{i=1}^I \frac{1}{4 t_i}\frac{1}{A_i(\mathbf{z})^2}\\
s.t. & \quad \mathbf{x}=\mathbf{z}.
\end{align}
Since the ADMM method is problem-specific, we leave the detailed analysis for the considered system model in the following sections.

\section{System model}
\begin{figure}[h]
	\centering
	\includegraphics[width=2.75in]{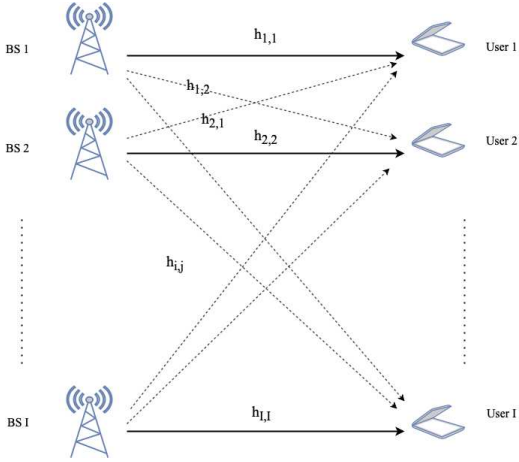}
	\caption{System model.}
	\label{figsys}
\end{figure}

Assume a multicell downlink scenario where the network has $I$ BSs and $I$ users as shown in Fig. \ref{figsys}. Each BS serves one user. All BSs share the same band, therefore introducing interference at the user side. The power gain from the $i$-th BS to the $j$-th user is denoted as $h_{i,j}$. 

In this system, $A_i(\mathbf{x})$ is interpreted as the rate for user $i$, $B_i(\mathbf{x})$ is the power consumption of BS $i$, and $\mathbf{x}$ is the vector of transmit power of all BSs. To avoid confusion, we will replace $\mathbf{x}$ by $\mathbf{p}$ in the following.

Denoting the $i$-th element in $\mathbf{p}$ as $p_i$ and the noise power as $\sigma^2$, the rate for user $i$ is expressed as 
\begin{equation}
A_i(\mathbf{p})=\log_2\left(1+\frac{h_{i,i}p_i}{\sum_{j\neq i}h_{j,i}p_j+\sigma^2}\right),
\end{equation}
and the power consumption for BS $i$ is expressed as 
\begin{equation}
B_i(\mathbf{p})=\phi_i p_i+Q_i,
\end{equation}
where $\phi_i$ is the inverse of amplifier efficiency and $Q_i$ is the circuit power of BS $i$. 
From these expressions, we know that 
$B_i(\mathbf{p})$ is linear; however, $A_i(\mathbf{p})$ is not concave.

The sum inverse EE (SIEE) minimization problem can be formulated as 
\begin{align}\label{ori}
\min_{\mathbf{p}\in C_p} \quad F_1(\mathbf{p})=\sum_{i=1}^I \frac{\phi_i p_i+Q_i}{\log_2\left(1+\frac{h_{i,i}p_i}{\sum_{j\neq i}h_{j,i}p_j+\sigma^2}\right)},
\end{align}
where $C_p$ is the domain of $\mathbf{p}$, which is the transmit power constraint.

\section{The solution of the minimization problem}
In this section, the solution of the proposed method is studied, which is divided into three steps. First, the problem is reformulated to deal with the non-concavity of the rate functions. Second, the reformulated problem is solved by the ADMM method. The optimization can be implemented distributedly and different parts of the problem can be solved individually. Third, the closed-form solutions are obtained thanks to the convexity of the reformulated problem and the ADMM method.

\subsection{Problem reformulation}
From Theorem \ref{theorem1}, we have the following equivalent problem:
\begin{align}
\min_{\mathbf{p}\in C_p, \mathbf{t}\in \mathbb{R}_{>0}^I} \quad F_2(\mathbf{p},\mathbf{t})=\sum_{i=1}^I t_i {B}_i(\mathbf{p})^2+\sum_{i=1}^I \frac{1}{4 t_i}\frac{1}{{A}_i(\mathbf{p})^2}.\label{biconvex}
\end{align}

To tackle with the non-concavity of ${A}_i(\mathbf{p})$, we introduce the following corollary, which is a direct result of Corollary 2 of paper \cite{yuwei}:

\begin{corollary}
	If $f_m$ is decreasing, then 
	\begin{equation}
	\min_{\mathbf{x}} \sum_m f_m\left(\frac{a_m(\mathbf{x})}{b_m(\mathbf{x})}\right)
	\end{equation} 
	is equivalent with 
	\begin{equation}
	\min_{\mathbf{x},\mathbf{y}} \sum_m f_m(2y_m \sqrt{a_m(\mathbf{x})}-y_m^2{b_m(\mathbf{x})}).
	\end{equation}
	Similarly with the update of $\mathbf{t}$, $\mathbf{y}$ can be updated by $y_m=\frac{\sqrt{a_m(\mathbf{x})}}{b_m(\mathbf{x})}$.
\end{corollary}

Therefore, to minimize $\sum_{i=1}^I \frac{1}{4t_i}\frac{1}{{A}_i(\mathbf{p})^2}$ is equivalent with 
\begin{equation}
\min_{\mathbf{p},\mathbf{y}} \sum_{i=1}^I \frac{1}{4t_i}\frac{1}{\hat{A}_i(\mathbf{p},y_i)^2},
\end{equation}
where
\begin{equation}
\hat{A}_i(\mathbf{p},y_i)=\log_2\left(1+2y_i\sqrt{h_{i,i}p_i}-y_i^2\left(\sum_{j\neq i}h_{j,i}p_j+\sigma^2\right)\right),
\end{equation}
which is biconcave w.r.t $\mathbf{p}$ and $y_i$. This means $\frac{1}{\hat{A}_i(\mathbf{p},y_i)^2}$ is biconvex due to Theorem \ref{theorem2}. 
Therefore, the following problem is a multi-convex problem:

\begin{align}
\max_{\mathbf{p}\in C_p, \mathbf{t}\in \mathbb{R}_{>0}^I, \mathbf{y}\in \mathbb{R}_{>0}^I} \quad F_3(\mathbf{p},\mathbf{t},\mathbf{y})=&\sum_{i=1}^I t_i {B}_i(\mathbf{p})^2\nonumber\\
&+\sum_{i=1}^I\frac{1}{4 t_i}\frac{1}{\hat{A}_i(\mathbf{p},y_i)^2},\label{biconvex2}
\end{align}
which is equivalent with problems \eqref{biconvex} and \eqref{ori}. A partial minimum can be efficiently found by alternate convex search, which is to optimize one variable while fixing others \cite{biconvex}.

\subsection{ADMM-based algorithm}
The updates of $\mathbf{t}$ and $\mathbf{y}$ are straightforward. Therefore, we focus on the update of $\mathbf{p}$ in the following.

For given $\mathbf{t},\mathbf{y}$, the problem is
\begin{align}
\max_{\mathbf{p}\in C_p} \quad F_3(\mathbf{p},\mathbf{t},\mathbf{y})=\sum_{i=1}^I t_i {B}_i(\mathbf{p})^2+\sum_{i=1}^I\frac{1}{4 t_i}\frac{1}{\hat{A}_i(\mathbf{p},y_i)^2}.\label{admm2}
\end{align}
Observing that ${B}_i(\mathbf{p})$ is only a function of $p_i$ and each $p_i$ has its own power constraint, it reminds us to decouple the terms of ${B}_i(\mathbf{p})$ and the terms of ${A}_i(\mathbf{p})$ to optimize ${B}_i(\mathbf{p})$ in a distributed manner. To this end, we use the ADMM method as stated in the following.

The augmented Lagrangian of \eqref{admm2} can be expressed as \cite{admm}
\begin{align}
L_\theta(\mathbf{p}, \mathbf{q}, \mathbf{u})=&\sum_{i=1}^I t_i {B}_i(\mathbf{p})^2+\frac{1}{4 t_i}\frac{1}{\hat{A}_i(\mathbf{q},y_i)^2}\nonumber\\&+\frac{\theta}{2}\|\mathbf{p}-\mathbf{q}+\mathbf{u}\|^2
+S(\mathbf{p}),
\end{align}
where $S(\mathbf{x})=0$ if $\mathbf{x}\in C_p$ and $S(\mathbf{x})=+\infty$ otherwise.

So the scaled form of ADMM is
\begin{align}
\mathbf{p}^{l+1}:=&\argmin_{\mathbf{p}\in C_p} \sum_{i=1}^I t_i {B}_i(\mathbf{p})^2+\frac{\theta}{2}\|\mathbf{p}-\mathbf{q}^{l}+\mathbf{u}^{l}\|^2\label{sub1}\\
\mathbf{q}^{l+1}:=&\argmin_{\mathbf{q}}\sum_{i=1}^I\frac{1}{4 t_i}\frac{1}{\hat{A}_i(\mathbf{q},y_i)^2}
+\frac{\theta}{2}\|\mathbf{p}^{l+1}-\mathbf{q}+\mathbf{u}^l\|^2\label{sub2}\\
\mathbf{u}^{l+1}:=&\mathbf{u}^l+\mathbf{p}^{l+1}-\mathbf{q}^{l+1}.\label{sub3}
\end{align}
The update in \eqref{sub3} is straightforward. Therefore, we will study how to solve \eqref{sub1} and \eqref{sub2} in the following.

\subsection{Closed-form solutions}
The Lagrangian of \eqref{sub1} can be written as 
\begin{align}
L(\mathbf{p},\{\lambda_i\},\{\beta_i\})=&\sum_{i=1}^I t_i {B}_i(\mathbf{p})^2+\frac{\theta}{2}\|\mathbf{p}-\mathbf{q}^{l}+\mathbf{u}^{l}\|^2\nonumber\\&+\sum_{i=1}^I \lambda_i(p_i-{P_i})-\beta_i p_i,
\end{align}
where $P_i$ is the power constraint for BS $i$. The KKT condition is
\begin{equation}
2t_i \phi_i(\phi_i p_i+Q_i)+\theta(p_i-q_i^l+u_i^l)+\lambda_i-\beta_i=0,
\end{equation}
which gives a closed-form solution as 
\begin{equation}
p_i=\left[\frac{\theta(q_i^l-u_i^l)-2t_i\phi_iQ_i}{2t_i\phi_i^2+\theta}\right]_0^{P_i}.
\end{equation}

Thanks to the ADMM method, the problem in 
\eqref{sub2} is now a constraint-free problem, as all constraints are on $\mathbf{p}$, not on $\mathbf{q}$. Because all $q_i$ are coupled, the optimization can only be implemented in a centralized manner. This unconstrained convex minimization can be solved by finding the stationary point, where
the derivative w.r.t. $q_i$ in \eqref{sub2} is 
\begin{align}
\sum_j-\frac{1}{2t_j}\frac{1}{\hat{A}_j(\mathbf{q},y_j)^3}\frac{\partial \hat{A}_j(\mathbf{q},y_j)}{\partial q_i}-\theta(p_i^{l+1}-q_i+u_i^l)=0.\label{newton}
\end{align}

Newton's method for system of equations can solve the equations, where the $i$-th equation is  $c_i(\mathbf{q})$, which is the opposite of the left-hand side of \eqref{newton} \cite{boyd}. Note that the formula to update the solution is 
\begin{equation}
\mathbf{q}_{n+1}=\mathbf{q}_n-J_C(\mathbf{q}_n)^{-1}C(\mathbf{q}_n),
\end{equation}
where $C(\mathbf{q})$ is $[c_1(\mathbf{q}),...,c_I(\mathbf{q})]^T$ and $J_C(\mathbf{q}_n)$ is the Jacobian matrix, whose $m,k$-th entry is $\frac{\partial c_m}{\partial x_k}$. To calculate the Jacobian matrix, some further manipulations and calculations are needed. We have
\begin{align}
c_i(\mathbf{q})=&\sum_j\frac{1}{2t_j}\frac{1}{\hat{A}_j(\mathbf{q},y_j)^3}\frac{\partial \hat{A}_j(\mathbf{q},y_j)}{\partial q_i}+\theta(p_i^{l+1}-q_i+u_i^l)\nonumber\\
=&\frac{y_i \sqrt{\frac{h_{i,i}}{q_i}}}{\ln 4 \cdot t_i \hat{A}_i(\mathbf{q},y_i)^3 g_i(\mathbf{q},y_i)}\nonumber\\
&-\sum_{j\neq i}\frac{y_j^2 h_{i,j}}{\ln 4 \cdot t_j \hat{A}_j(\mathbf{q},y_j)^3 g_j(\mathbf{q},y_j)}\nonumber\\
&+\theta(p_i^{l+1}-q_i+u_i^l),
\end{align}
where $g_i(\mathbf{q},y_i)=1+2y_i\sqrt{h_{i,i}q_i}-y_i^2\left(\sum_{j\neq i}h_{j,i}q_j+\sigma^2\right)$.

By defining 
\begin{equation}
D_i=\frac{3}{\ln 2\cdot\hat{A}_i(\mathbf{q},y_i)^4 g_i(\mathbf{q},y_i)^2 }+\frac{1}{\hat{A}_i(\mathbf{q},y_i)^3 g_i(\mathbf{q},y_i)^2},
\end{equation}
we have

	\begin{align}
	\frac{\partial c_i(\mathbf{q})}{\partial q_i}=&-\frac{y_i^2h_{i,i}}{\ln 4\cdot t_i q_i}D_i-\frac{y_i}{\ln 16\cdot t_i\hat{A}_i(\mathbf{q},y_i)^3 g_i(\mathbf{q},y_i)}\sqrt{\frac{h_{i,i}}{q_i^3}}\nonumber\\
	&-\sum_{j\neq i}\frac{y_j^4 h_{i,j}^2}{\ln 4 \cdot t_j}D_j-\theta.
\end{align}

If $m\neq i$, then

	\begin{align}
	\frac{\partial c_i(\mathbf{q})}{\partial q_m}=&\frac{y_i^3 h_{m,i}\sqrt{\frac{h_{i,i}}{q_i}}}{\ln 4 \cdot t_i}D_i
	\nonumber\\&-\sum_{j\neq m,j\neq i}\frac{y_j^4 h_{m,j}h_{i,j}}{\ln 4 \cdot t_j}D_j
	+\frac{y_m^3 h_{i,m}\sqrt{\frac{h_{m,m}}{p_m}}}{\ln 4\cdot t_m}D_m.
	\end{align}

Thus, by using Newton's method, the solution for \eqref{sub2} is found. The closed-form expression for each iteration in Newton's method has been obtained.

\section{Algorithms}
In this section, based on the analysis above, we propose and summarize the alternate convex search to solve the problem in Algorithm \ref{al1}.
We begin the algorithm by initializing the newly introduced $\mathbf{t}$ and $\mathbf{y}$ for reformulations, and $\mathbf{u}$ and $\mathbf{q}$ for the ADMM method. Convergence is guaranteed since the problem is a multi-convex problem.

%

\begin{algorithm}[H]
	\caption{Alternate convex search}
	\begin{algorithmic}\label{al1}
		\STATE Initialize $\delta_1$ and $\delta_2$; initialize $\mathbf{t}^{(0)}$ and $\mathbf{y}^{(0)}$; initialize $\mathbf{u}^{0}$ and $\mathbf{q}^{0}$; set $n=0$
		\WHILE {true}
		\STATE $l=0$; $n=n+1$
		\WHILE {true}
		\STATE Update $\mathbf{p}^{l+1}$ using \eqref{sub1}
		\STATE Update $\mathbf{q}^{l+1}$ using \eqref{sub2}
		\STATE $y_i^{(n)}=\frac{\sqrt{h_{i,i}q_i}}{\sum_{j\neq i}h_{j,i}q_j+\sigma^2}$ for each $i$
		\STATE Update $\mathbf{u}^{l+1}$ using \eqref{sub3}
		\STATE $l=l+1$
		\IF {$|\mathbf{p}^{l+1}-\mathbf{q}^{l+1}|<\delta_1$}
		\STATE Break;
		\ENDIF
		\ENDWHILE
		\STATE $t_i^{(n)}=\frac{1}{2\hat{A}_i(\mathbf{p},y_i)B_i(\mathbf{p})}$ for each $i$
		\IF {$|\mathbf{t}^{(n)}-\mathbf{t}^{(n-1)}|<\delta_2$}
		\STATE Break;
		\ENDIF
		\ENDWHILE
	\end{algorithmic}
\end{algorithm}

\begin{figure}[!t]
	\centering
	\includegraphics[width=3.5in]{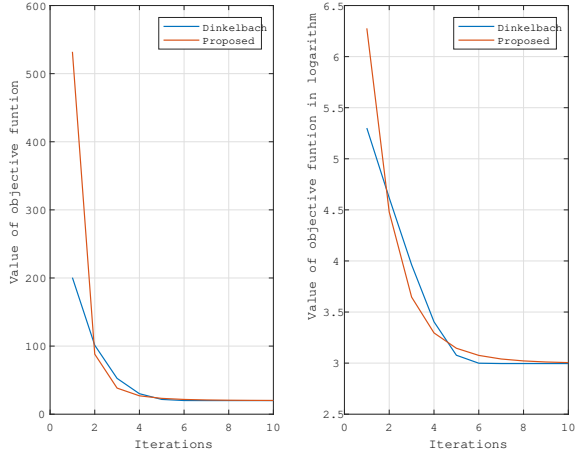}
	\caption{Comparison between Dinkelbach and the proposed algorithm.}
	\label{f1}
\end{figure}

\begin{figure}[h]
	\centering
	\includegraphics[width=3.5in]{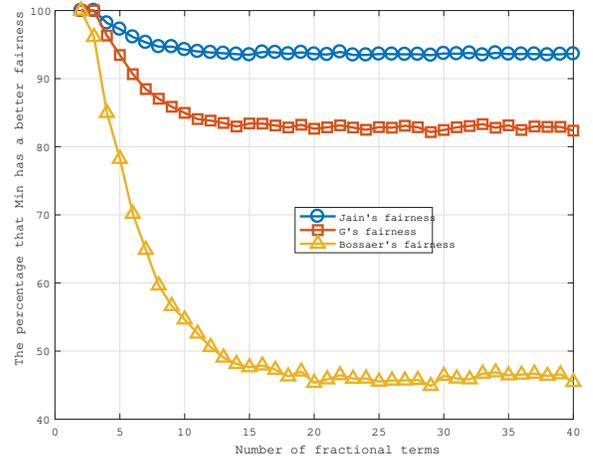}
	\caption{Fairness comparison: small range (5)}
	\label{f2}
\end{figure}
\begin{figure}[h]
	\centering
	\includegraphics[width=3.5in]{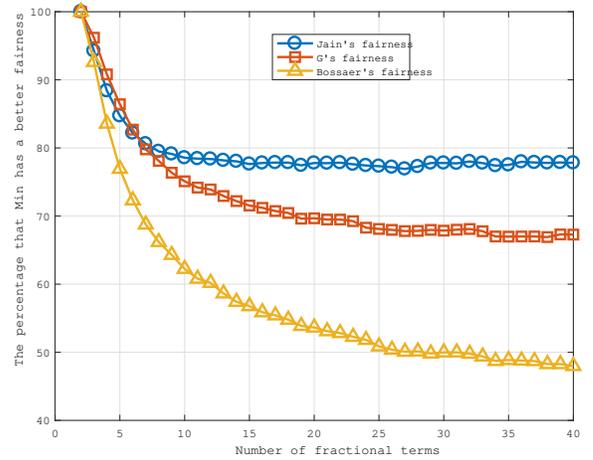}
	\caption{Fairness comparison: medium range (10).}
	\label{f3}
\end{figure}

\begin{figure}[h]
	\centering
	\includegraphics[width=3.5in]{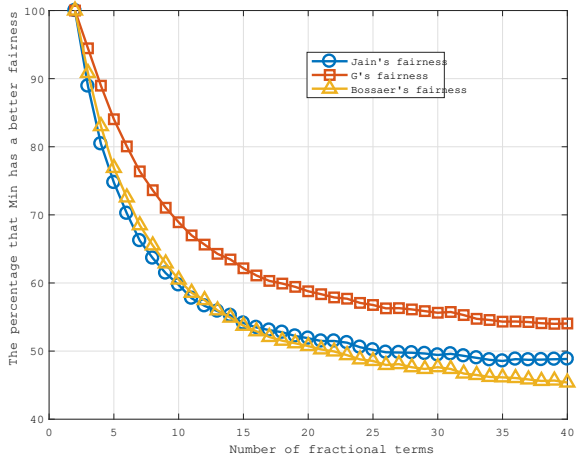}
	\caption{Fairness comparison: large range (50).}
	\label{f4}
\end{figure}

\section{Numerical and simulation results}
In this section, we will illustrate the theoretical results by means of numerical results and simulation results.

The numerical results in  Fig. \ref{f1} show that the proposed fraction transform has convergence speed similar to that of Dinkelbach's algorithm for a fractional programming.
In particular, we minimize $\frac{x^2+100}{x}$ using Dinkelbach algorithm and the proposed algorithm respectively. Both algorithms obtain roughly the optimal solution within five iterations.

\begin{figure}[h]
	\centering
	\includegraphics[width=3.5in]{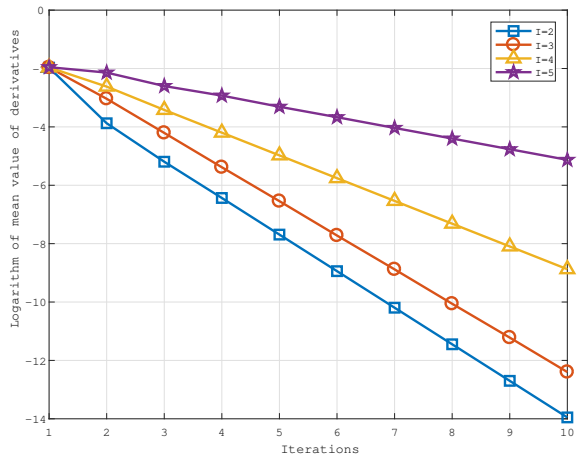}
	\caption{Convergence speed of Newton's method for different $I$.}
	\label{f5}
\end{figure}

\begin{figure}[h]
	\centering
	\includegraphics[width=3.5in]{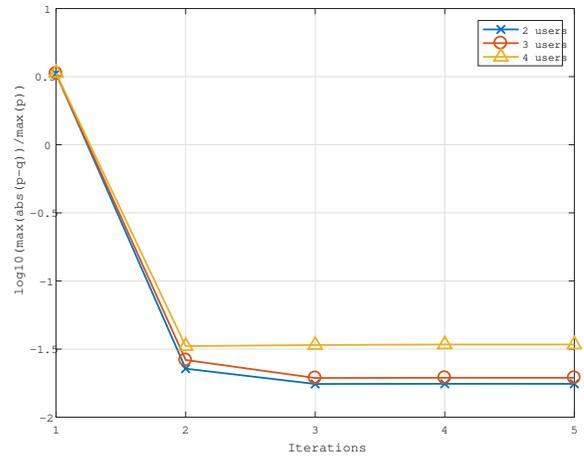}
	\caption{The convergence of the primal residual of the ADMM in the proposed algorithm.}
	\label{f9}
\end{figure}

\begin{figure}[h]
	\centering
	\includegraphics[width=3.5in]{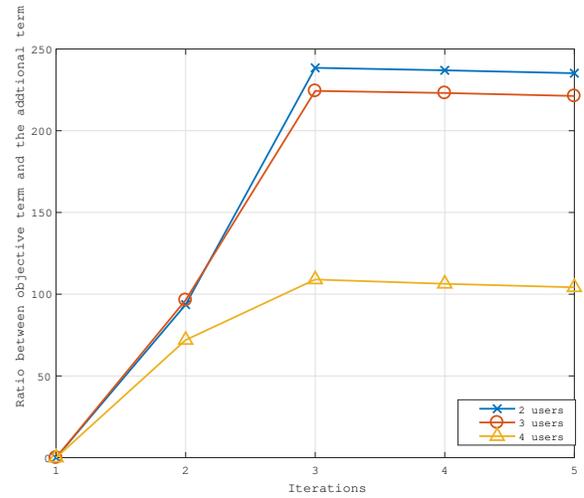}
	\caption{The ratio between the value of objective function and the value of the additional term.}
	\label{f10}
\end{figure}

\begin{figure}[h]
	\centering
	\includegraphics[width=3.5in]{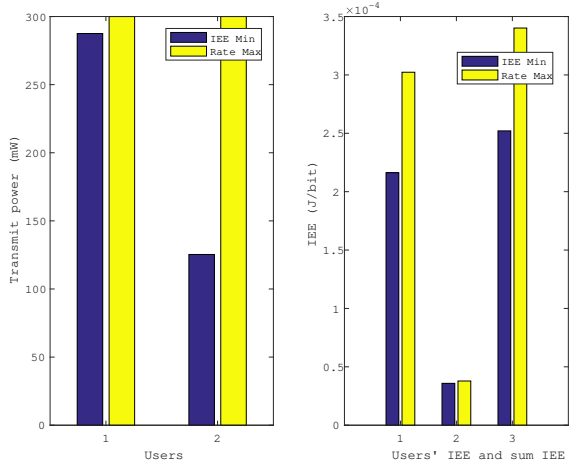}
	\caption{Comparison of transmit power, individual IEE, and sum IEE for 2 users.}
	\label{f6}
\end{figure}

\begin{figure}[h]
	\centering
	\includegraphics[width=3.5in]{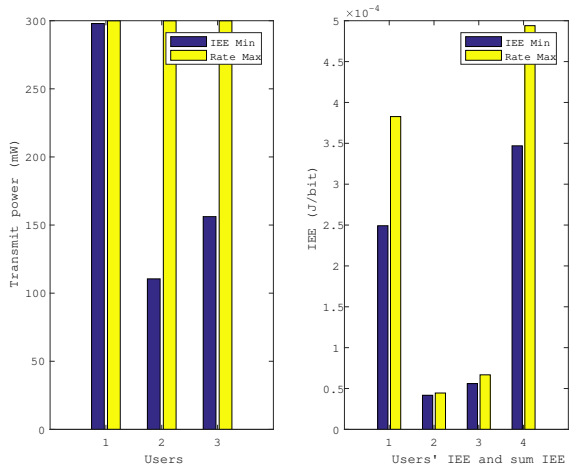}
	\caption{Comparison of transmit power, individual IEE, and sum IEE for 3 users.}
	\label{f7}
\end{figure}

The numerical results in Fig. \ref{f2}, Fig. \ref{f3}, and Fig. \ref{f4} show the fairness comparison between SIMin and SMax. Three criteria of fairness are considered: Jain's fairness, G's fairness, and Bossaer's fairness. Each term in the sum maximization problem (like $x$ and $y$ in \eqref{fair1}) is chosen from a random variable uniform between $1$ and a maximum value equal to respectively $5$, $10$, $50$ in different figures. The percentages that SIMin have better fairness are plotted versus the number of fractional terms. 'A better fairness' means a larger value of a certain fairness criterion. It is observed that, for a small range of $5$, SIMin is better than Jain's fairness and G's fairness with high probability. With Bossaer's fairness, it is better from 2 terms to 12 terms. As the dynamic range increases, the possibility that SIMin has a better fairness decreases. However, it can be observed that, when the number of terms is less than around 15, SIMin is better than the SMax for all ranges. Please note that a range of $50$, which, under the context of EE, means one user's EE is $50$ times the other one's, is already quite large. One can refer to Fig. \ref{f6} and Fig. \ref{f7} as examples, which show that the highest EE is around $8$ times the lowest one. 
Therefore, we can conclude that for two terms, it is mathematically proven that SIMin is always better than SMax. For less than 15 terms, these numerical results show that the fairness of SIMin is most of the time better than the one of SMax.

Next we illustrate our simulation results based on the system model. The system parameters are set as follows: the bandwidth for each subcarrier is set to $B = 10$KHz, $Q_i=0.5$mW. The power constraints are $P_i=0.3$mW. We also select the following values for the channel modeling: $n=3.5$,  $G_0=-(G_1+M_l)=-70$dB, where $G_1 = 30$dB is the gain factor at d = 1m and $M_l = 40$dB \cite{zijian}. The noise power spectral density is set to $N_0=-170$dBm/Hz, the noise figure to $N_f = 10$dB/Hz and the inverse of amplifier efficiency is chosen to be $\phi_i=2.5$.

The most complex procedure is Newton's method to solve \eqref{sub2}.
Fig. \ref{f5} shows the comparison of convergence speed of Newton's method for different $I$. It is observed that the algorithm converges within few iterations. The convergence for large values of $I$ is also sufficiently fast.

The efficiency of ADMM method also needs to be validated. 

The primal residual should converge to a small value. This is illustrated in Fig. \ref{f9}. We observe that, for various numbers of BSs, the primal residual is much smaller than $1$, meaning $\mathbf{p}$ and $\mathbf{q}$ are close enough in their respective subproblems.

The value of the original objective function should be sufficiently larger than the additional term introduced for the convergence of $\mathbf{p}$ and $\mathbf{q}$. This is validated in Fig. \ref{f10}. We observe that, for various numbers of BSs, the value of the original objective function is more than 100 times larger than the additional term.

In Fig. \ref{f6} and Fig. \ref{f7}, we compare the proposed optimization with rate maximization for 2 and 3 users respectively. The performance improvement in terms of SIEE from rate maximization to SIEE minimization is significant. It is assumed that user 1 has a weak channel and user 2 and 3 have stronger channels. As shown in the figure, this improvement comes mainly from the user with a worse channel (user 1 in these figures). This confirms the fairness improvement by the SIEE minimization, which reduces the difference of the values of IEE among users. This improvement is achieved by reducing the transmit power of users with better channels, which is an essential observation for interference channels: reducing transmit power of users with good channels does not influence much of its own EE, but improves the EE of users with weak channels.

\section{Conclusion}
In this paper, a framework of solving SFMin problems is proposed and a SIEE minimization problem is solved for multiple BS systems.  Two new vector variables are introduced to reformulate the original problem into a multi-convex problem. The ADMM is used to further simplify the problem to obtain closed-form solutions. Numerical results confirm the fairness improvement of SIMin compared with SMax. Simulation results show that the algorithm convergences fast and the ADMM method is efficient. The EE performance outperforms the conventional rate maximization.

\section*{Acknowledgment}
This work was supported by FNRS (Fonds National de la recherche scientifique) under EOS project Number 30452698. The authors would like to thank UCL for funding the ARC SWIPT project.

\end{document}